\documentclass[fleqn,10pt]{wlscirep}
\newcommand{\degree}{\ensuremath{^\circ}}

\title{Computational insights and the observation of SiC nanograin assembly: towards 2D silicon carbide}

\author[1,*]{Toma Susi}
\author[1,2]{Viera Sk\'akalov\'a}
\author[1]{Andreas Mittelberger}
\author[3]{Peter Kotrusz}
\author[3]{Martin Hulman}
\author[1]{Timothy J. Pennycook}
\author[1]{Clemens Mangler}
\author[1]{Jani Kotakoski}
\author[1,*]{Jannik C. Meyer}

\affil[1]{University of Vienna, Faculty of Physics, Boltzmanngasse 5, 1090 Vienna, Austria}
\affil[2]{Slovak University of Technology (STU), Center for Nanodiagnostics, Vazovova 5, 812 43 Bratislava, Slovakia}
\affil[3]{Danubia NanoTech, Ilkovicova 3, 841 04 Bratislava, Slovakia}

\affil[*]{toma.susi@univie.ac.at \& jannik.meyer@univie.ac.at}

\begin{abstract}
While an increasing number of two-dimensional (2D) materials, including graphene and silicene, have already been realized, others have only been predicted. An interesting example is the two-dimensional form of silicon carbide (2D-SiC). Here, we present an observation of atomically thin and hexagonally bonded nanosized grains of SiC assembling temporarily in graphene oxide pores during an atomic resolution scanning transmission electron microscopy experiment. Even though these small grains do not fully represent the bulk crystal, simulations indicate that their electronic structure already approaches that of 2D-SiC. This is predicted to be flat, but some doubts have remained regarding the preference of Si for \textit{sp}$^{3}$ hybridization. Exploring a number of corrugated morphologies, we find completely flat 2D-SiC to have the lowest energy. We further compute its phonon dispersion, with a Raman-active transverse optical mode, and estimate the core level binding energies. Finally, we study the chemical reactivity of 2D-SiC, suggesting it is like silicene unstable against molecular absorption or interlayer linking. Nonetheless, it can form stable van der Waals-bonded bilayers with either graphene or hexagonal boron nitride, promising to further enrich the family of two-dimensional materials once bulk synthesis is achieved.
\end{abstract}
\begin{document}

\flushbottom
\maketitle
\thispagestyle{empty}

\section*{Introduction}

In the wake of the 2004 discovery of graphene, the single-atom thin form of hexagonal carbon~\cite{Geim07NM}, two-dimensional (2D) materials have attracted increasing attention. They can be divided into two classes: inherently layered materials bound by van der Waals interactions, including hexagonal boron nitride (hBN)~\cite{Watanabe04NM}, phosphorene~\cite{Li14NN} and transition metal dichalgonenides such as molybdenum disulphide~\cite{Wang12NN}; and those with non-planar covalent bonding in their bulk form. An important class of the latter consists of the remaining group-IV elements, namely Si, Ge, Sn and Pb. The first of these, composed on Si and named silicene~\cite{Takeda94PRB,Cahangirov09PRL}, has been synthesized on silver substrates~\cite{Aufray10APL,Vogt12PRL}, and further fabricated into transistors~\cite{Tao15NN}. Unlike graphene, silicene exhibits a chair-like distortion of the hexagonal rings, resulting in out-of-plane corrugation. Like graphene, charge carriers in silicene show a Dirac dispersion at the Fermi level~\cite{Cahangirov09PRL}, although a small gap does open due to the structural distortion~\cite{Tao15NN}.

In addition to lattices of either pure C or Si, mixed stoichiometries are possible for 2D forms of silicon carbide (2D-Si$_{x}$C$_{1-x}$)~\cite{Zhou13NL,Gao13JACS}. Although the \textit{s}$^2$\textit{p}$^2$ valence shell structure of Si is similar to C, its greater covalent bonding distance in most crystals inhibits \textit{p}--\textit{p} overlap, leading to \textit{sp}$^{3}$ hybridization. Bulk SiC exists in over 200 crystalline forms~\cite{Cheung12}, some with three-dimensional hexagonal crystal structures. For the isoatomic 2D form Si$_{0.5}$C$_{0.5}$ (which we will simply call 2D-SiC), a planar structure identical to graphene but with a bond distance of 1.77-1.79~\AA---compared to 1.425~\AA~for graphene, 1.89~\AA~for bulk SiC, and 2.33~\AA~for bulk silicon---and a large band gap (2.5--2.6~eV) have been predicted~\cite{Sahin09PRB,Hsueh11PRB,Lin13JMCC}. A recent cluster expansion study explored the space of possible C:Si mixings, finding the lowest formation energy for the isoatomic stoichiometry~\cite{Shi15AN}.

In terms of experimental efforts, Lin et al. were recently able to synthesize flakes of quasi-2D SiC and SiC$_2$ via high-temperature thermochemical substitution reactions of exfoliated graphene with Si powder~\cite{Lin15JPCC}. Although these flakes were not atomically thin ($<$10 nm in thickness) and thus closer to bulk polytypes than true 2D-SiC, the authors did observe some differences in electronic structure indicative of confinement. Even more recently, Chabi et al. were able to use a similar method to push the average thickness down to 2-3 nm, but did not provide evidence of their electronic properties~\cite{Chabi16N}. Despite such efforts and the explosion of interest in two-dimensional materials, no truly 2D form of SiC has yet been realized.

We present here atomic resolution scanning transmission electron microscopy (STEM) observations of nanosized grains of SiC, found to assemble in the pores of graphene oxide (GO)~\cite{Dikin07N}. GO is useful here for two reasons: its disordered structure contains nanometer-sized holes (as in disordered graphene~\cite{Robertson15AN}), and the synthesis by-products remaining even after purification provide an ample source of mobile C and Si adatoms. The observed patches were stable for tens of seconds under the intense 60 keV electron irradiation, allowing us to capture high quality images of the atomic configurations. The observed bonding and the measured annular dark field detector intensities precisely match a quantitative image simulation based on a density functional theory (DFT) model. In contrast to earlier observations of pure Si cluster dynamics in a graphene nanopore~\cite{Lee13NC}, our grains contain a similar number of C and Si atoms, predominantly in an alternating arrangement.

We do not claim to have synthesized 2D-SiC, nor suggest that grain assembly is a practical route to the bulk material. Nonetheless, it is remarkable to directly observe the formation of a hexagonal SiC lattice. Our simulations further indicate that the electronic structure in the largest patches does already resemble that of the bulk form, for which we compute many important characteristics to guide further experimental efforts.

\section*{Results and discussion}
\subsection*{Samples and microscopy}

\begin{figure}
\centering
\includegraphics[width=15.2cm]{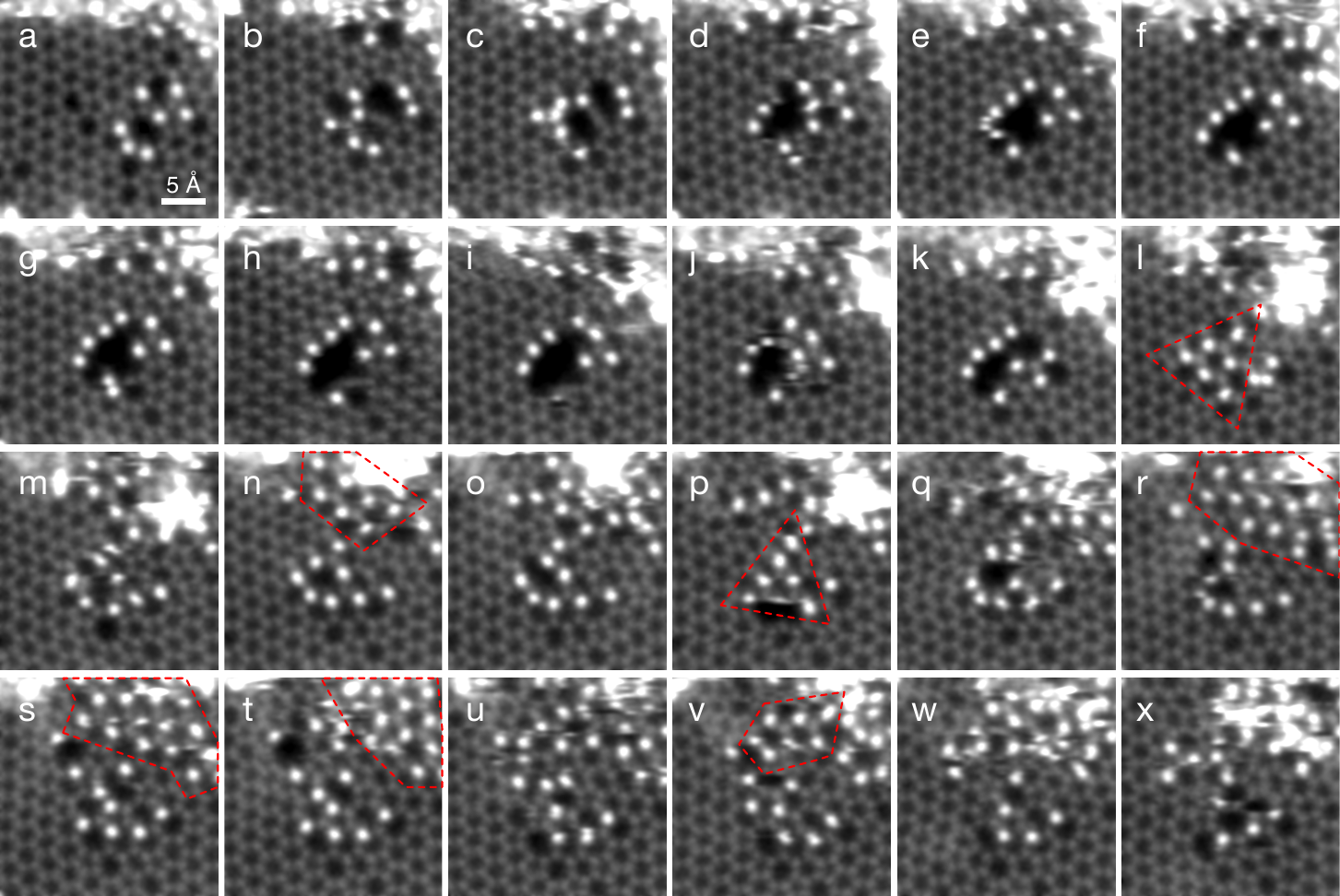}  
\caption{Observed time series of the assembly of atomically thin SiC nanograins. The overlaid red dashed lines indicate the approximate locations of the SiC regions. The triangular patch in panel l was used as the basis for elemental identification.\label{Fig1}}
\end{figure}

Our graphene oxide synthesis has been described in detail previously~\cite{Skakalova14C} (see also Methods). In our transferred samples, a good coverage of single-layer GO flakes was found on the support grid. To study their morphology, we conducted electron microscopy in a Nion UltraSTEM100 scanning transmission electron microscope operated at 60 kV. While continuously imaging a clean monolayer area of a flake, we were surprised to observe the conversion of a disordered area surrounding a small pore into a hexagonal lattice of alternating heavier and lighter atoms (Fig.~\ref{Fig1}, time between panels $\sim$23 s, panels smoothed for clarity with a 0.1 \AA\ Gaussian kernel). Heavier atoms appear brighter in images recorded with an annular dark field detector, since greater Coulomb scattering of the probe electrons occurs from the nuclei of atoms with more protons (Z-contrast~\cite{Krivanek10N}). A particularly well resolved structure in panel l exhibited a triangular crystalline patch of 3$\times$3 units, whose structure could be precisely deduced from the image. By comparison with simulations (shown further below) we can identify all lighter atoms as carbon, and the heavier ones as silicon. The lighter atoms inside the patch appear slightly brighter than the carbon atoms in the graphene lattice, which is a well known effect of probe tails in presence of the heavier neighbouring atoms~\cite{Krivanek10N}. 

Under the intense electron irradiation, this patch was not stable for long, but later in the series another larger patch formed near the top right corner of the view (Figure~\ref{Fig1} panels l, n, p, r-t, v). Thus, even though the irradiation continuously perturbs the atomic structure, the dynamics are not fully random and there seems to be an energetic tendency towards this periodic arrangement of atoms. The assembly of the SiC lattice may thus be considered as the result of beam-driven sampling of the dynamical potential energy landscape. Interestingly, the lines of Si atoms in different crystallites, or in the same one at different times, do not have the same angles with respect to the graphene lattice. This suggests that the pores merely act as suitable containing spaces~\cite{Zhao14S}, and do not have a strong role in directing the assembly. An additional example, starting with heavier atoms saturating the reactive edge of a pore~\cite{Zan12NL,Chen16AN}, is shown as Supplementary Fig.~S1, reminiscent of the reknitting of holes in graphene~\cite{Zan12NL}.

\subsection*{Structure identification}
Using the experimental image (Figure \ref{Fig1}l) as a starting point, we created a simplified symmetrical model structure of six Si atoms embedded in a 9$\times$9 supercell of graphene and relaxed its atomic structure via DFT using the \textsc{gpaw} package~\cite{Enkovaara2010} (Methods). We chose to omit the differently bonded Si atoms from the edge of the patch to keep the periodic unit cell manageable for high accuracy calculations, resulting in a model of 150 atoms in total (with six atomic substitutions and 12 missing C atoms). The bond lengths between the Si and C in the relaxed structure varied between 1.78 and 1.87 \AA , depending on the atom pair. To create a larger structure for image simulations, we repeated the cell periodically and cropped a square of 27.0$\times$25.4 \AA~(627 atoms) surrounding the patch. In addition, we created a primitive 2-atom unit cell for 2D-SiC. By relaxing the structure and optimizing the cell size using the stress tensor in plane-wave mode (Methods), we found a planar ground state with a Si--C bond length of 1.792~\AA. Supercells of this were used to calculate a number of properties of the material and compare them to those of the primary bulk forms (Table \ref{Table1}), which also helped confirm the accuracy of our simulations.

To identify the atoms in Fig. \ref{Fig1}l, we used the QSTEM software package~\cite{Koch2002} to find a quantitative match of intensities between experimental and simulated images. As we know that most atoms in our field of view are carbon, we could use the graphene lattice contrast as reference and subtract a background value measured in vacuum in a large hole with the same imaging conditions from the raw data prior to measuring the intensities. The image contrast is influenced by lens aberrations (including chromatic aberration ~\cite{Kuramochi09UM}), thermal diffuse scattering ~\cite{Forbes11UM}, finite source size and, importantly, the annular dark field detector angles. All these can be addressed by the QSTEM simulation and were set to values representing our experimental setup (Methods).

\begin{figure}
\centering\includegraphics[width=7cm]{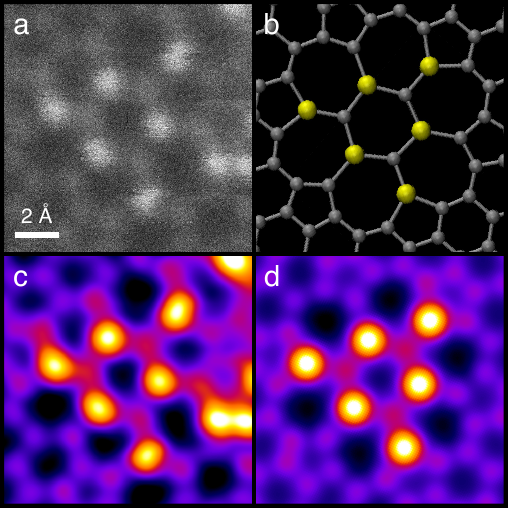}
\caption{Elemental identification of the atoms within the SiC nanograin. (a) Crop of an unprocessed MAADF-STEM image. (b) Simplified atomic SiC model. (c) The experimental image with noise removed by a Gaussian blur (sigma = 0.28 \AA), with higher intensities coloured towards white. (d) Crop of a quantitative image simulation of the SiC model (see text).\label{Fig2}}
\end{figure}

The image thus simulated (Figure \ref{Fig2}d) leads to a Si/C intensity ratio of 2.15 (average for all 6 Si atoms and 10 C atoms away from the SiC patch), with an increased intensity on the C atoms next to Si due to the probe tails~\cite{Krivanek10N} (although the model structure differs from the experimental structure at its edges, this does not affect the intensities of the central atoms). From the experimental image, we measured a Si/C intensity ratio of 2.17, matching the simulation with an error of only 1\%. No other impurity element provides a similar ratio. Two-dimensional silica~\cite{huang12NL}, on the other hand, has a much larger lattice, and oxygen atoms would not form three bonds or be beam-stable~\cite{Tararan16CM}. Thus the investigated structure can only be SiC, which is not surprising since Si is frequently found in graphene samples~\cite{Susi14PRL,Chen16AN}, especially in graphene oxide prepared via wet chemistry (although its origin is still unknown). We could also verify the presence of Si as a major contamination in this sample by electron energy loss spectroscopy (Supplementary Fig. S2), but individual atoms were too mobile to reliably confirm their identity by spectroscopy.

\subsection*{Electronic structure of the nanograins}
The electronic band structure of bulk 2D-SiC has already been extensively discussed in the literature~\cite{Sahin09PRB,Lin13JMCC}. In our case, the electronic band gap was estimated by converging the band structure of the primitive 2-atom cell up to 8 unoccupied bands (Methods), yielding a gap of 2.58~eV (predicted to be as high as 4.42 eV due to unusually strong excitonic effects included at the \textsc{G$^0$W$^0$} level of theory~\cite{Hsueh11PRB}). The overall band structure (Supplementary Fig.~S3) is in good agreement with earlier simulations~\cite{Hsueh11PRB,Lin13JMCC}.

\begin{figure}
\centering\includegraphics[width=11.5cm]{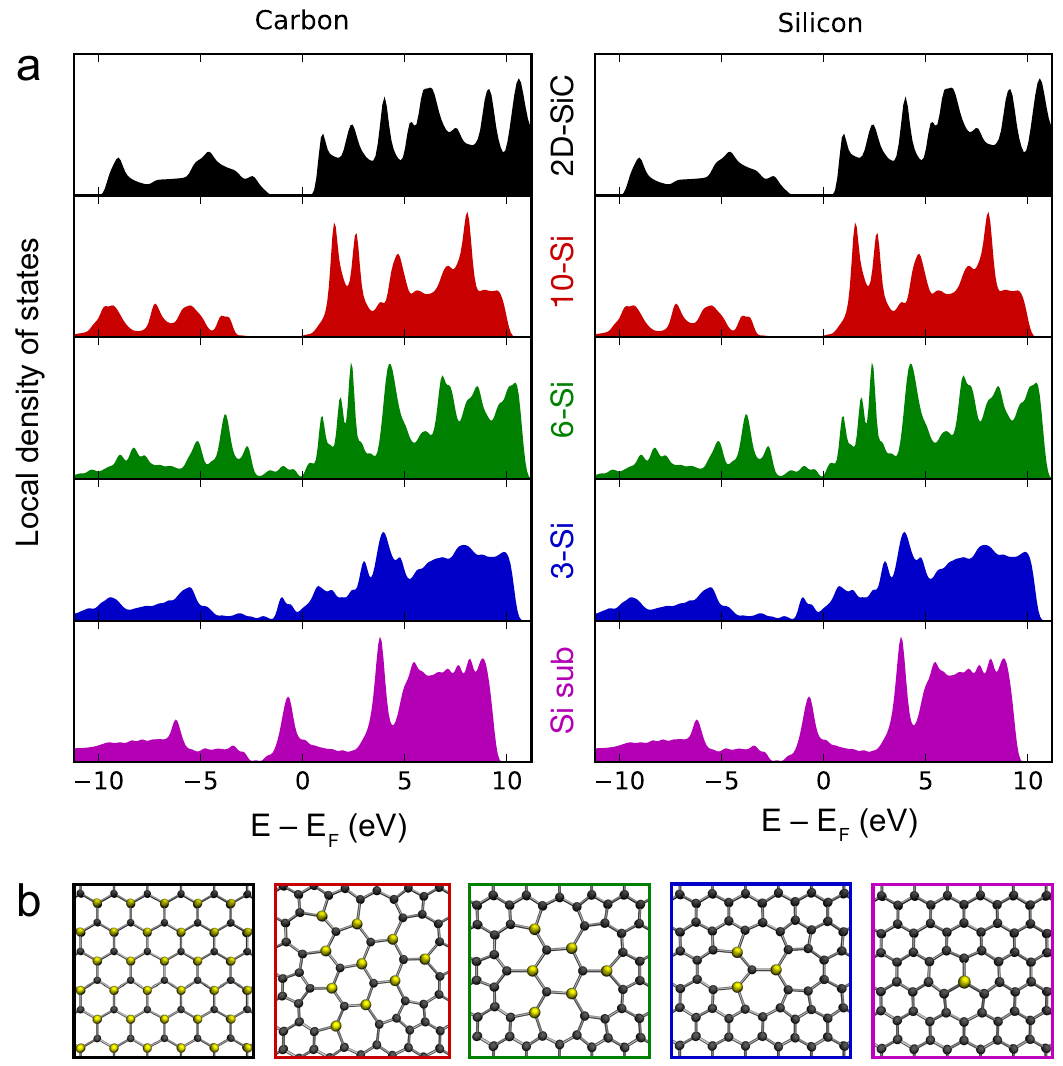}
\caption{Local densities of electronic states (LDOS) for SiC nanograins of different sizes and for 2D-SiC. a) LDOSes projected onto Si- or C-centred Wigner-Seitz cells of 2D-SiC compared to those of the SiC-like atoms in smaller patches embedded into graphene (the legend identifies the size of the calculated SiC grain and applies to both panels). b) Structure models used for the LDOS projections (C atoms shown in black, Si in yellow; frame colours correspond to the line colours in a).\label{Fig3}}
\end{figure}

Using this band structure as a starting point to assess whether the nanograins resemble the bulk form in their properties, we calculated the Wigner-Seitz local densities of state (LDOS, Fig.~\ref{Fig3}a) of the SiC-like atoms in a single Si substitution and triangular patches of SiC with 3, 6 and 10 Si atoms (Fig.~\ref{Fig3}b), and compared them to 2D-SiC. A clear trend towards the bulk electronic structure can be observed despite the distortions in the atomic structure due to the embedding strain (which is known~\cite{Lin13JMCC} to affect the band gap of 2D-SiC), with the 10-Si patch exhibiting a clear band gap. Thus it appears that despite their small size, the largest patches we observed could already be considered as nanosized grains of the material. Note also that since we are comparing the local densities of states inside the patch, the angle with respect to the lattice or the precise arrangement of atoms outside the patch should not affect this result.

\subsection*{The planarity and hybridization of 2D-SiC}
Aside from the clear hexagonal bonding, the projected bond lengths in our experimental images are consistent with a planar 2D-SiC structure. In the literature, simulated 2D-SiC has typically been characterized as flat~\cite{Sahin09PRB,Hsueh11PRB,Lin13JMCC,Shi15AN}, but it has not been clear if there have been enough atoms in the chosen unit cells to allow for corrugation, and whether or not the chosen cell sizes have been imposing strain that prevents buckling. The flatness is somewhat surprising considering the propensity of Si to prefer \textit{sp}$^{3}$ bonding, which leads to puckering in the ground state of silicene~\cite{Cahangirov09PRL}. Furthermore, the lowest energy structure of an analogous material---two-dimensional phosphorus carbide (2D-PC)---was very recently predicted to be highly corrugated~\cite{Guan16NL}.

To address this, we ran calculations in rectangular 4 and 8-atom unit cells, starting from different degrees of corrugation and cell sizes, including structures with alternating C--C and Si--Si bonds (analogous to the proposed ground state of 2D-PC). In all cases, the total energy of the relaxed structure was minimized for Si--C bonding (consistent with Shi et al.~\cite{Shi15AN}), and lowest for an entirely planar structure (Supplementary Fig. S4). Thus, while there may be competition between \textit{sp}$^{2}$ hybridization preferred by C in its planar form and \textit{sp}$^{3}$ preferred by Si, the ground state of 2D-SiC is indeed planar. 

Bader analysis~\cite{Tang09JoPCM} further reveals that the Si--C bond in 2D-SiC is rather polarized~\cite{Lin13JMCC,Shi15AN}, with Si donating almost 1.2 electrons to its three C neighbours. To understand the bond hybridization in more detail, we projected the Kohn-Sham orbitals of a 48-atom rectangular supercell to the maximally localized Wannier orbitals~\cite{Thygesen05PRL} of the \textit{sp}$^2$-bonded carbosilane analogue of ethene (SiH$_2$CH$_2$), with its Si and C atoms fixed to the locations corresponding to a single Si--C bond. The resulting projector overlaps were close to unity, indicating that these \textit{sp}$^2$ molecular orbitals provide a good representation of the bond.

\subsection*{Phonon band structure and cohesive energy}
\begin{figure}
\centering\includegraphics[width=8.7cm]{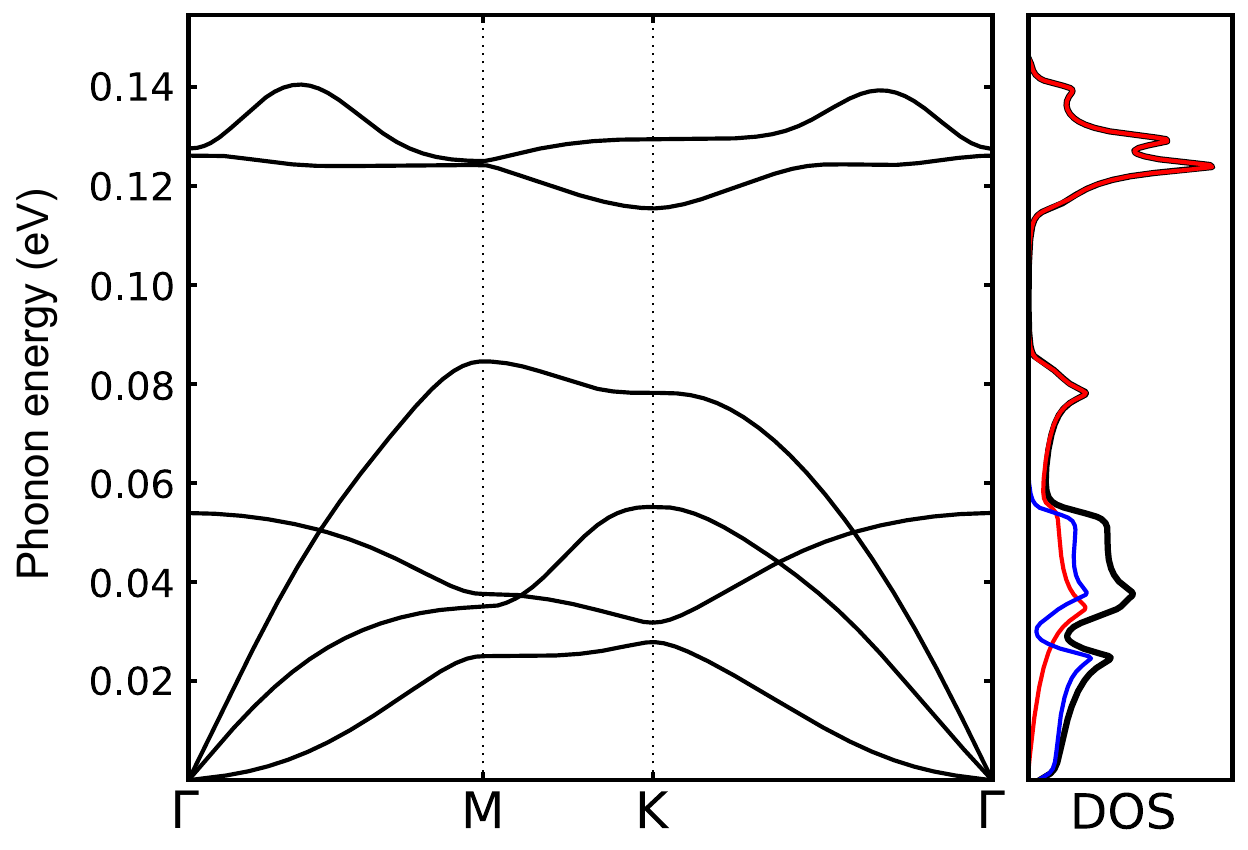}
\caption{The phonon band structure of 2D-SiC and the corresponding density of states (in-plane and out-of-plane components are shown in red and blue, respectively).\label{Fig4}}
\end{figure}

We then calculated the phonon band structure of 2D-SiC through its dynamical matrix, estimated by displacing each primitive cell atom by a 0.08~\AA\ displacement in the three Cartesian directions and calculating via DFT the forces on all other atoms in a 7$\times$7 supercell (the so-called 'frozen phonon' approximation; Methods). Unlike that of an earlier calculation~\cite{Sahin09PRB}, the resulting phonon band structure  (Fig.~\ref{Fig4}) contains no imaginary frequencies, demonstrating the stability of the material~\cite{Shi15AN}. The energy of the Raman-active transverse optical branch at $\Gamma$ is 127~meV, anticipating a G-band-like feature at 1024~cm$^{-1}$. It thus seems clear that fully planar 2D-SiC indeed is stable, and while its cohesive energy (PBE functional) is 0.50 eV/atom lower than that of 3C-SiC (the main bulk SiC polymorphs have very similar energies \cite{Kackell94PRB}), that difference is smaller than that between silicene and monocrystalline Si (0.64 eV/at.).

\subsection*{Other properties}
Further properties of 2D-SiC can be computationally predicted. In terms of electron irradiation stability, Si is too heavy to be displaced from the SiC structure at acceleration voltages below 100 kV. We calculated the displacement threshold energy $T_{d}$ for the C atom via DFT molecular dynamics (MD), described in detail previously~\cite{Kotakoski11PRB,Susi12AN,Kotakoski12AN,Susi14PRL,Susi16NC}. In brief, we estimated $T_{d}$ by increasing the starting out-of-plane kinetic energy of a selected C atom until it escaped the structure during the course of an MD simulation. For the structure shown in Fig. \ref{Fig2}, the energy required to displace a C atom from the SiC patch is approximately 13.25~eV. Although this is higher than what can be transferred to a static nucleus, it is low enough that atomic vibrations can enable displacements~\cite{Meyer12PRL,Susi14PRL,Susi16NC} and bond rotations~\cite{Kotakoski11PRB}. For a C atom in bulk 2D-SiC ($7\times7$ supercell), the threshold is instead 15.75~eV, leading to a negligible displacement probability by 60~keV electrons at room temperature. Thus a macroscopic flake of 2D-SiC should prove rather stable for low-voltage microscopy. We also calculated its bulk modulus by uniaxially straining the optimal 2D-SiC unit cell and finding the minimum of the resulting total energy curve, resulting in 98.3~GPa.

Finally, we estimated the C~1\textit{s} and Si~2\textit{p} core level binding energies of 2D-SiC via delta Kohn--Sham ($\mathrm{\Delta}$KS) total energy differences including an explicit core-hole~\cite{Ljungberg11JESRP,Susi15PRB} (Methods). The C~1\textit{s} energy was calculated at 283.265~eV and the Si~2\textit{p} at 101.074~eV. Although the absolute values are sensitive to the accuracy of the description of core-hole screening, the energy separation C~1\textit{s} -- Si~2\textit{p} of 182.19~eV should characterize 2D-SiC well.

\begin{table}[b]
  \small
    \caption{Comparison of 2D-SiC properties we calculated to those of the major bulk polytypes reported in the literature (from Ref.~\citenum{IoffeSiC} unless otherwise indicated).\label{Table1}}
   \begin{tabular*}{0.6\textwidth}{@{\extracolsep{\fill}}lllll}
\hline
    Polytype & 2D-SiC & 6H ($\alpha$) & 4H & 3C ($\beta$)\\
\hline
    Symmetry& hexagonal & hexagonal & hexagonal & cubic \\
       In-plane lattice constant (\AA)& 3.104 & 3.0810 & 3.0730 & 4.3596\\
       Si-C bond length (\AA) & 1.792 & 1.89 & 1.89 & 1.89 \\
    Bandgap (eV) & 2.58 & 3.05 & 3.23 & 2.36\\
            Bulk modulus (GPa) & 98.3 & 220 & 220 & 250 \\
           Optical phonon energy (meV) & 127 & 102.8 & 104.2 & 104.2 \\
           C 1\textit{s} -- Si 2\textit{p} (eV) & 182.19 & 181.9~\cite{Binner01JMSL} & 182.3~\cite{Johansson96SS} &  182.17~\cite{Parrill91SS} \\
        \hline
  \end{tabular*}
\end{table}
\normalsize

\subsection*{Reactivity and bilayers}
Considering the large charge transfer and unconventional hybridization of the flat structure, we suspected that 2D-SiC might be chemically reactive. To study this computationally, we first hydrogenated a monolayer of 2D-SiC with atomic H (in analogy to graphane~\cite{Sofo07PRB}). The H preferentially bonds with C, with a formation energy of 0.79~eV with respect to the chemical potential of H$_2$. However, a second H bonds to Si on the opposite side of the plane, resulting in a highly corrugated structure when the cell is allowed to relax, bringing the formation energy down to -1.23~eV. Thus, similar to silicene and phosphorene~\cite{Molle13AFM,Wood14NL}, 2D-SiC likely is unstable in air.

Finally, to estimate the stability of 2D-SiC in bilayers, we completed several calculations using a van der Waals exchange correlation functional~\cite{Cooper10PRB} (in the plane-wave mode, see Methods). First, we initialized a simulation with two 2D-SiC layers 4~\AA~apart in each of the five possible stacking orders (in analogy to hBN~\cite{Constantinescu13PRL}). Although AB, AB', A'B and AA stacking resulted in stable bilayers, the lowest energy (by 34, 41, 54 and 55 meV/atom, respectively) is obtained for AA' stacking where the Si are located over the C and vice versa (Fig. \ref{Fig5}a). Minimizing the forces brings the two layers to within 2.32~\AA~of each other, inducing a slight corrugation of the planes. The residual stresses of the unit cell indicate that its size prevents the structure from reaching its (three-dimensional) ground state, and analysis of the all-electron density between the layers clearly indicates covalent bonding (Fig. \ref{Fig5}b), in almost perfect analogy to bilayer silicene~\cite{Padilha15JPCC}. This confirms that no van der Waals bonded layered equivalent of 2D-SiC can exist, in agreement with its absence among the known phases of bulk SiC. However, when the other layer is instead either graphene (a 5$\times$5 supercell of graphene has only a 0.5\% lattice mismatch to a 4$\times$4 supercell of 2D-SiC) or hBN (-1.4\% mismatch), the equilibrium distances are $\sim$3.5~\AA~(possibly slightly affected by the lattice mismatch in our simulation cell) with binding energies of $\sim$56~meV per atom (finite-difference mode, see Methods), both consistent with van der Waals bonding. This suggests that encapsulation could be used to protect 2D-SiC from the atmosphere without seriously affecting its properties.

\begin{figure}
\centering\includegraphics[width=9cm]{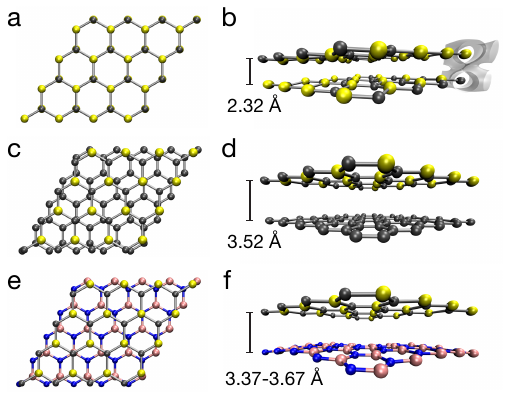}
\caption{Calculated equilibrium structures of bilayers of 2D-SiC with itself (a-b), graphene (c-d) and hexagonal boron nitride (hBN, e-f). (a-b) Two layers of 2D-SiC in AA' stacking spontaneously bond covalently (all-electron charge density isosurface shown in the corner of the cell in b), resulting in an interlayer distance of 2.32 \AA. When the other layer is graphene (c-d) or hBN (e-f), the equilibrium distances and binding energies are typical for van der Waals bonding. (Note that the resulting hBN structure is slightly buckled due to lattice mismatch in the simulation unit cell.)\label{Fig5}}
\end{figure}

\section*{Conclusions}
In conclusion, our atomic resolution scanning transmission electron microscopy observations provide the first direct experimental indication that a two-dimensional form of silicon carbide may exist. Pores in graphene oxide act as templating spaces, with the electron beam effectively providing an energy input so that the Si-C configuration space is explored. During this process, mobile adatoms of C and Si provide a chemical source. As revealed by extensive simulations, the ground state of bulk 2D-SiC is indeed completely planar, with \textit{sp}$^2$ hybridization of the Si--C bond. The large charge transfer from Si to C and the preference of Si for \textit{sp}$^{3}$ hybridization render the layer chemically reactive and unstable in bilayers, similar to several other 2D materials. However, our simulations indicate that bilayers of 2D-SiC with either graphene or hexagonal boron nitride are stable, making it a promising candidate for incorporation into layered van der Waals heterostructures~\cite{Geim13N}.

\section*{Methods}

\subsection*{Sample preparation}

Graphite powder (purity 99.9995\%, 2-15 $\mu$m flakes, Alfa Aesar) was mixed into sulphuric acid, and then potassium permanganate and sodium nitrate added portion-wise. For the oxidation, water was added and the reaction mixture heated to 98\degree C~for 3 weeks. Terminating the reaction was followed by filtering, washing, and drying. To exfoliate the resulting graphite oxide powder into single-layer flakes, it was mixed with deionized water, vigorously stirred for 24 h, followed by bath sonication for 3 h, tip sonication for 30 min, and finally bath sonication for a further 1 h. To prepare the TEM samples, a Au support grid covered with a holey carbon film (Quantifoil\textregistered) was dipped into a water-based dispersion for 1 min and then rinsed in isopropanol and dried in air~\cite{Meyer08NL}.

\subsection*{Electron microscopy}

The Nion UltraSTEM100 scanning transmission electron microscope was operated at 60 kV in near-ultrahigh vacuum ($\sim$2$\times10^{-7}$ Pa). The beam current during the experiments was a few tens of pA, corresponding to a dose rate of approximately $1\times10^{7}$ e$^{-}/$\AA$^{2}$s. The beam convergence semiangle was 35~mrad and the semi-angular range of the medium-angle annular dark field (MAADF) detector was 60--80~mrad.

\subsection*{Density functional theory}

The larger cell calculations were conducted using the GPAW finite-difference mode with a 0.18~\AA~grid spacing and 3$\times$3$\times$1 Monkhorst-Pack \textbf{k}-points. For the plane-wave calculations, we used a cutoff energy of 600 eV (increased to 700 eV for the band structure) and 45$\times$45$\times$1 \textbf{k}-points. The Perdew-Burke-Ernzerhof (PBE) functional was used to describe exchange and correlation, except for the bilayer simulations where we used the C09 van der Waals functional~\cite{Cooper10PRB}. 

For calculating the phonon band structure, we used instead the local density approximation (LDA) and a $\Gamma$-centred k-point mesh of 42$\times$42$\times$1 was used to sample the Brillouin zone. A fine computational grid spacing of 0.16~\AA~alongside strict convergence criteria for the structural relaxation (forces $<10^{-5}$ eV/\AA~per atom) and the self-consistency cycle (change in eigenstates $<10^{-13}$ eV$^{2}$ per electron) ensured accurate forces.

For the core level calculations, an extra electron was introduced into the valence band to ensure charge neutrality, and supercells up to 9$\times$9 in size used to confirm that spurious interactions between periodic images of the core hole were minimized. Although spin-orbit interaction was not included in the Si~2\textit{p} calculation, its binding energy can be taken to correspond to the 2\textit{p}$_{3/2}$ level and a splitting of 0.63~eV inferred from theory.

\subsection*{Image simulation}

Our QSTEM parameters were: chromatic aberration coefficient of 1 mm with an energy spread of 0.3 eV; spherical aberration coefficient of 1 $\mu$m; thermal diffuse scattering included via frozen phonon modelling with a temperature of 300 K; additional instabilities (such as sample vibration) simulated by blurring the resulting image (Gaussian kernel with a sigma of 0.39 \AA); and the medium-angle annular dark-field detector angle range set to the experimental range of 60--80 mrad. Shot noise was removed from the filtered experimental image of Fig. \ref{Fig2}c by blurring it with a Gaussian kernel (sigma of 0.28 \AA).

\subsection*{Data availability}
All data generated or analysed during this study are included in this published article and its Supplementary Information.

\section*{Acknowledgements}

We thank Michael Walter and Miguel Caro for useful discussions. T.S. acknowledges funding from the Austrian Science Fund (FWF) via project P 28322-N36 and the Vienna Scientific Cluster for computational resources. V.S. was supported by the FWF via project AI0234411/21, and through projects VEGA 1/1004/15 and UVP, OPVaV-2011/4.2/01-PN. A.M. and J.C.M. acknowledge funding by the FWF project P25721-N20 and the European Research Council Grant No. 336453-PICOMAT. T.J.P. was supported by the European Union's Horizon 2020 research and innovation programme under the Marie Skłodowska-Curie grant agreement No. 655760 -- DIGIPHASE, and J.K. by the Wiener Wissenschafts\mbox{-,} Forschungs- und Technologiefonds (WWTF) via project MA14-009.

\section*{Author contributions statement}

T.S. conducted the DFT simulations and drafted the manuscript. V.S. conceived of the study, conducted electron microscopy, and supervised the sample synthesis. A.M. simulated the STEM images. P.K. and M.H. synthesized the samples. T.J.P. and C.M. participated in electron microscopy. J.K. and J.C.M. supervised the study. All authors reviewed the manuscript. 

\section*{Additional information}

\textbf{Supplementary information} accompanies this paper at http://www.nature.com/ scientificreports.
\\
\textbf{Competing financial interests} The authors declare no competing financial interests.

\begin{figure}
  \renewcommand\figurename{Supplementary Figure}
  \setcounter{figure}{0}
\centering\includegraphics[scale=1.3]{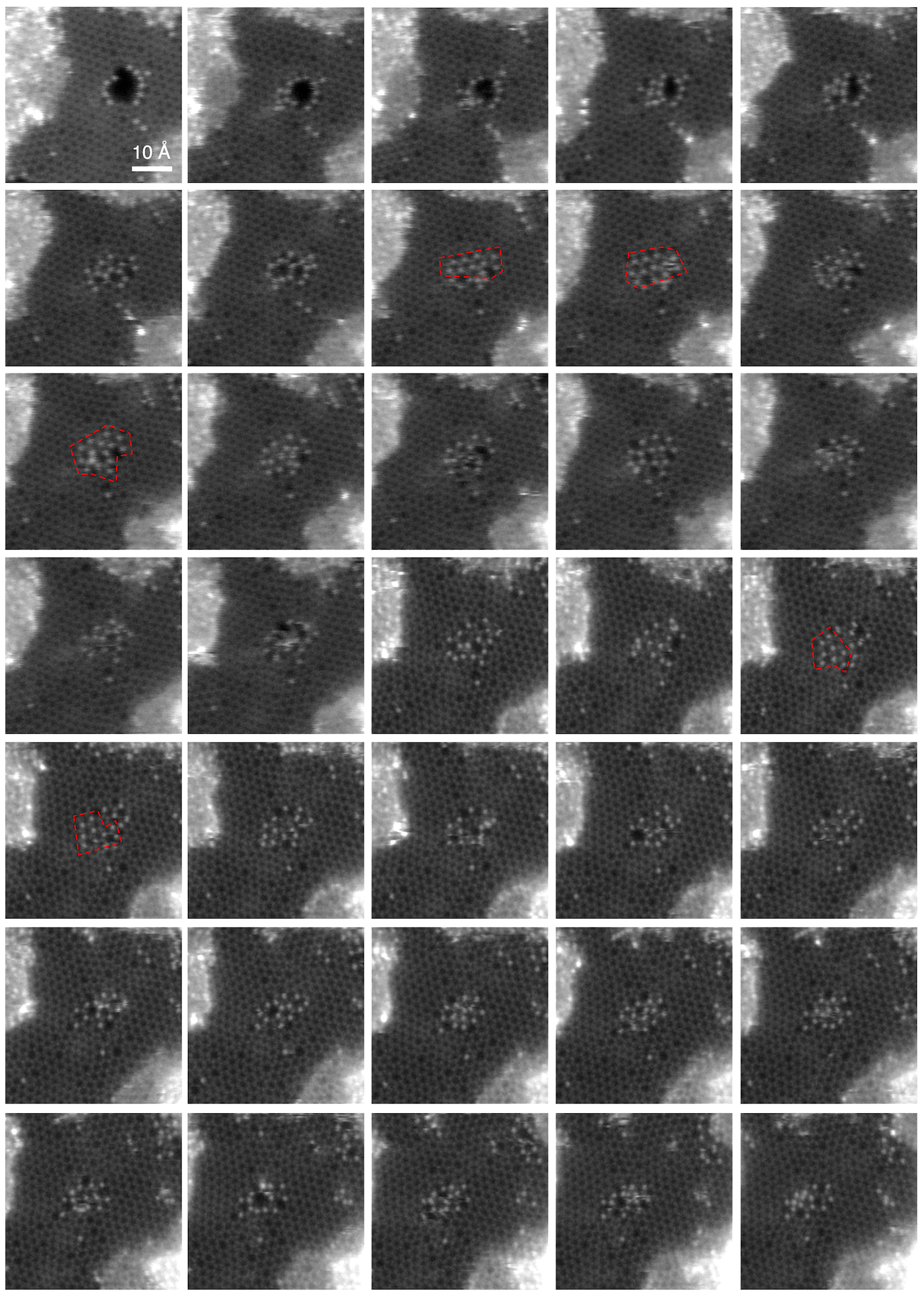}
\caption{An additional time series of SiC grain assembly, with particularly well resolved grains outlined with red dashed lines.\label{SI1}}
\end{figure}

\begin{figure}
  \renewcommand\figurename{Supplementary Figure}
  \setcounter{figure}{1}
\centering\includegraphics[scale=1.6]{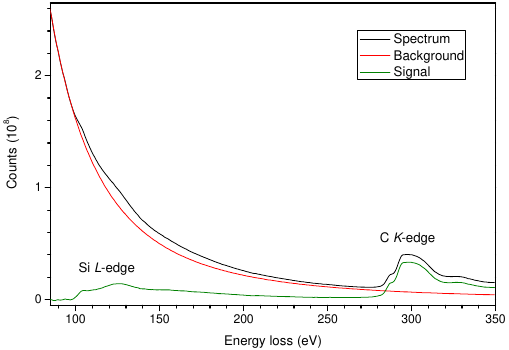}
\caption{An electron energy loss spectrum recorded over the graphene oxide lattice (0.73 eV/px dispersion). The black line is the original spectrum, red is a background fit, and green the resulting signal. Only the C \textit{K}-edge starting at $\sim$280 eV and the Si \textit{L}-edge starting at $\sim$99 eV are present in the spectrum.\label{SI2}}
\end{figure}

\begin{figure}
  \renewcommand\figurename{Supplementary Figure}
  \setcounter{figure}{2}
\centering\includegraphics[scale=0.7]{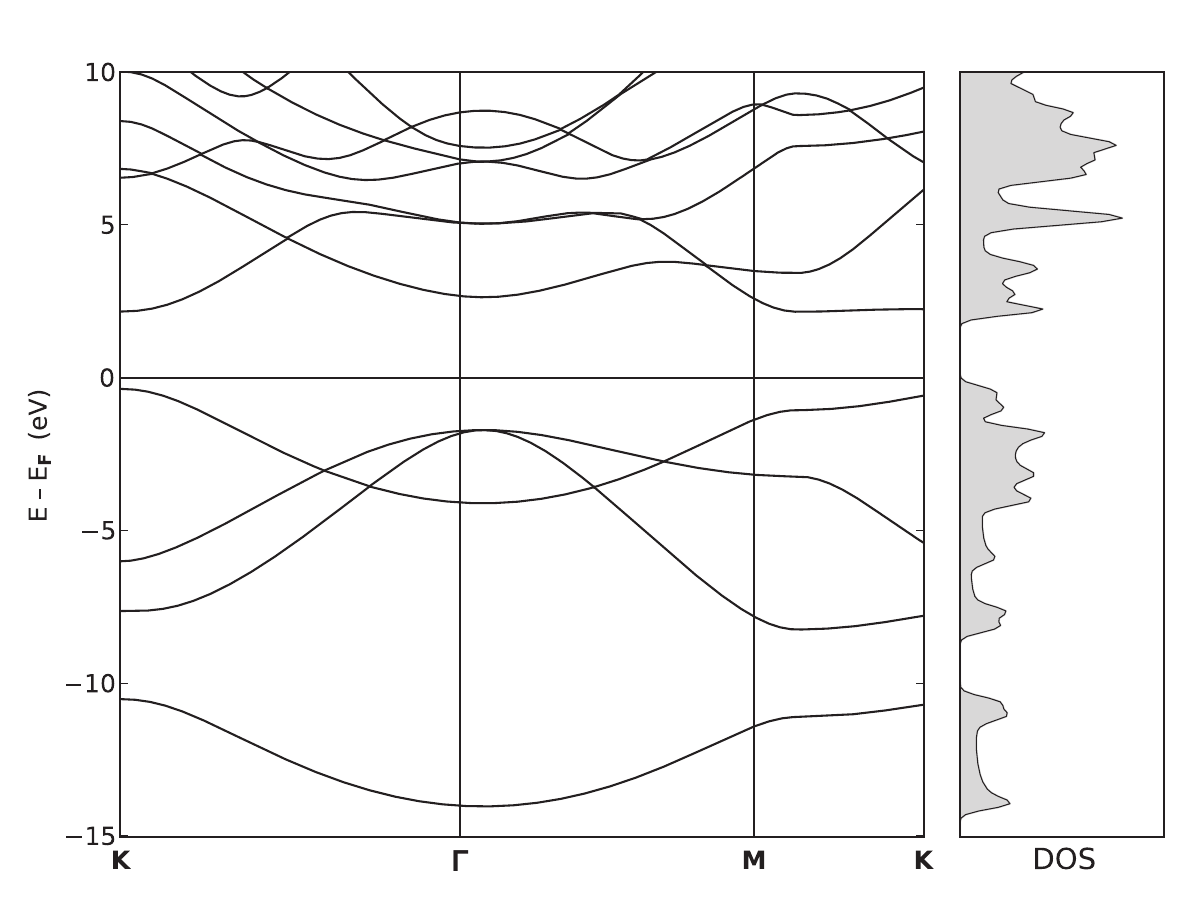}
\caption{The near-gap electronic bandstructure of 2D-SiC calculated with the PBE functional, and the corresponding density of states.\label{SI3}}
\end{figure}

\begin{figure}
  \renewcommand\figurename{Supplementary Figure}
  \setcounter{figure}{3}
\centering\includegraphics[scale=0.55]{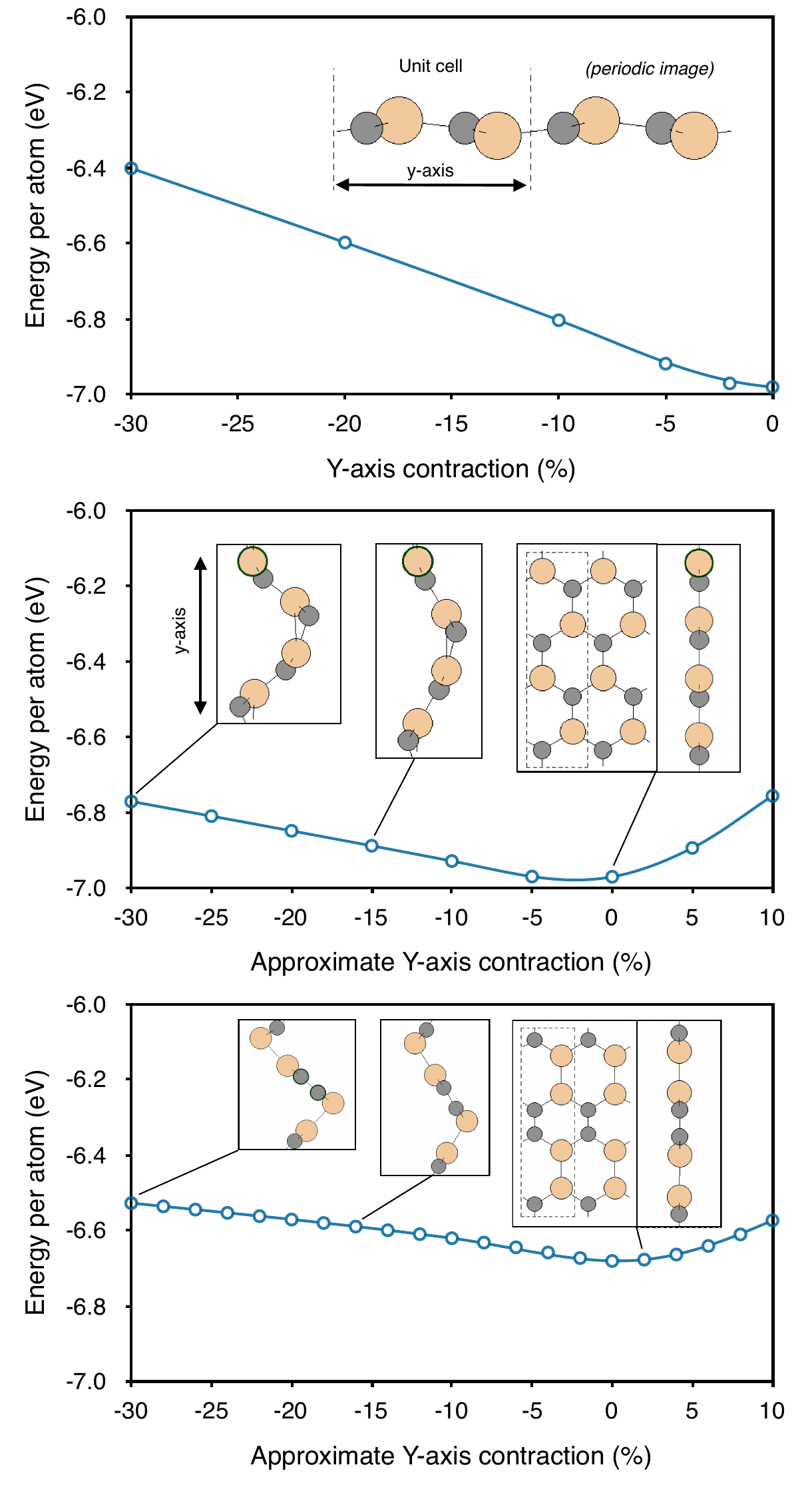}
\caption{Energies of distorted 2D-SiC calculated with density functional theory, demonstrating that the fully flat structure is the ground state. Relaxation of the cell in the perpendicular direction (not shown) does not appreciably change the energies.\label{SI4}}
\end{figure}


\begin{thebibliography}{10}
\expandafter\ifx\csname url\endcsname\relax
  \def\url#1{\texttt{#1}}\fi
\expandafter\ifx\csname urlprefix\endcsname\relax\def\urlprefix{URL }\fi
\expandafter\ifx\csname doiprefix\endcsname\relax\def\doiprefix{DOI }\fi
\providecommand{\bibinfo}[2]{#2}
\providecommand{\eprint}[2][]{\url{#2}}

\bibitem{Geim07NM}
\bibinfo{author}{Geim, A.~K.} \& \bibinfo{author}{Novoselov, K.~S.}
\newblock \bibinfo{title}{The rise of graphene}.
\newblock \emph{\bibinfo{journal}{Nat. Mater.}} \textbf{\bibinfo{volume}{6}},
  \bibinfo{pages}{183--191} (\bibinfo{year}{2007}).

\bibitem{Watanabe04NM}
\bibinfo{author}{Watanabe, K.}, \bibinfo{author}{Taniguchi, T.} \&
  \bibinfo{author}{Kanda, H.}
\newblock \bibinfo{title}{Direct-bandgap properties and evidence for
  ultraviolet lasing of hexagonal boron nitride single crystal}.
\newblock \emph{\bibinfo{journal}{Nat. Mater.}} \textbf{\bibinfo{volume}{3}},
  \bibinfo{pages}{404} (\bibinfo{year}{2004}).

\bibitem{Li14NN}
\bibinfo{author}{Li, L.} \emph{et~al.}
\newblock \bibinfo{title}{Black phosphorus field-effect transistors}.
\newblock \emph{\bibinfo{journal}{Nature Nanotechnology}}
  \textbf{\bibinfo{volume}{9}}, \bibinfo{pages}{372--377}
  (\bibinfo{year}{2014}).

\bibitem{Wang12NN}
\bibinfo{author}{Wang, Q.~H.}, \bibinfo{author}{Kalantar-Zadeh, K.},
  \bibinfo{author}{Kis, A.}, \bibinfo{author}{Coleman, J.~N.} \&
  \bibinfo{author}{Strano, M.~S.}
\newblock \bibinfo{title}{Electronics and optoelectronics of two-dimensional
  transition metal dichalcogenides}.
\newblock \emph{\bibinfo{journal}{Nature Nanotechnology}}
  \textbf{\bibinfo{volume}{7}}, \bibinfo{pages}{699--712}
  (\bibinfo{year}{2012}).

\bibitem{Takeda94PRB}
\bibinfo{author}{Takeda, K.} \& \bibinfo{author}{Shiraishi, K.}
\newblock \bibinfo{title}{Theoretical possibility of stage corrugation in {S}i
  and {G}e analogs of graphite}.
\newblock \emph{\bibinfo{journal}{Phys. Rev. B}} \textbf{\bibinfo{volume}{50}},
  \bibinfo{pages}{14916--14922} (\bibinfo{year}{1994}).

\bibitem{Cahangirov09PRL}
\bibinfo{author}{Cahangirov, S.}, \bibinfo{author}{Topsakal, M.},
  \bibinfo{author}{Akt\"urk, E.}, \bibinfo{author}{\ifmmode~\mbox{\c{S}}\else
  \c{S}\fi{}ahin, H.} \& \bibinfo{author}{Ciraci, S.}
\newblock \bibinfo{title}{Two- and one-dimensional honeycomb structures of
  silicon and germanium}.
\newblock \emph{\bibinfo{journal}{Phys. Rev. Lett.}}
  \textbf{\bibinfo{volume}{102}}, \bibinfo{pages}{236804}
  (\bibinfo{year}{2009}).

\bibitem{Aufray10APL}
\bibinfo{author}{Aufray, B.} \emph{et~al.}
\newblock \bibinfo{title}{Graphene-like silicon nanoribbons on {A}g(110): A
  possible formation of silicene}.
\newblock \emph{\bibinfo{journal}{Applied Physics Letters}}
  \textbf{\bibinfo{volume}{96}} (\bibinfo{year}{2010}).

\bibitem{Vogt12PRL}
\bibinfo{author}{Vogt, P.} \emph{et~al.}
\newblock \bibinfo{title}{Silicene: Compelling experimental evidence for
  graphenelike two-dimensional silicon}.
\newblock \emph{\bibinfo{journal}{Phys. Rev. Lett.}}
  \textbf{\bibinfo{volume}{108}}, \bibinfo{pages}{155501}
  (\bibinfo{year}{2012}).

\bibitem{Tao15NN}
\bibinfo{author}{Tao, L.} \emph{et~al.}
\newblock \bibinfo{title}{Silicene field-effect transistors operating at room
  temperature}.
\newblock \emph{\bibinfo{journal}{Nature Nanotechnology}}
  \textbf{\bibinfo{volume}{10}}, \bibinfo{pages}{227--231}
  (\bibinfo{year}{2015}).

\bibitem{Zhou13NL}
\bibinfo{author}{Zhou, L.-J.}, \bibinfo{author}{Zhang, Y.-F.} \&
  \bibinfo{author}{Wu, L.-M.}
\newblock \bibinfo{title}{Si{C}$_2$ siligraphene and nanotubes: Novel donor
  materials in excitonic solar cells}.
\newblock \emph{\bibinfo{journal}{Nano Letters}} \textbf{\bibinfo{volume}{13}},
  \bibinfo{pages}{5431--5436} (\bibinfo{year}{2013}).

\bibitem{Gao13JACS}
\bibinfo{author}{Gao, G.}, \bibinfo{author}{Ashcroft, N.~W.} \&
  \bibinfo{author}{Hoffmann, R.}
\newblock \bibinfo{title}{The unusual and the expected in the {S}i/{C} phase
  diagram}.
\newblock \emph{\bibinfo{journal}{Journal of the American Chemical Society}}
  \textbf{\bibinfo{volume}{135}}, \bibinfo{pages}{11651--11656}
  (\bibinfo{year}{2013}).

\bibitem{Cheung12}
\bibinfo{author}{Cheung, R.}
\newblock \emph{\bibinfo{title}{Introduction to Silicon Carbide
  Microelectromechanical Systems (MEMS)}}, chap.~\bibinfo{chapter}{1},
  \bibinfo{pages}{1--17} (\bibinfo{publisher}{Imperial College Press, London},
  \bibinfo{year}{2012}).

\bibitem{Sahin09PRB}
\bibinfo{author}{\ifmmode~\mbox{\c{S}}\else \c{S}\fi{}ahin, H.} \emph{et~al.}
\newblock \bibinfo{title}{Monolayer honeycomb structures of group-{IV} elements
  and {III}-{V} binary compounds: First-principles calculations}.
\newblock \emph{\bibinfo{journal}{Phys. Rev. B}} \textbf{\bibinfo{volume}{80}},
  \bibinfo{pages}{155453} (\bibinfo{year}{2009}).

\bibitem{Hsueh11PRB}
\bibinfo{author}{Hsueh, H.~C.}, \bibinfo{author}{Guo, G.~Y.} \&
  \bibinfo{author}{Louie, S.~G.}
\newblock \bibinfo{title}{Excitonic effects in the optical properties of a
  {SiC} sheet and nanotubes}.
\newblock \emph{\bibinfo{journal}{Phys. Rev. B}} \textbf{\bibinfo{volume}{84}},
  \bibinfo{pages}{085404} (\bibinfo{year}{2011}).

\bibitem{Lin13JMCC}
\bibinfo{author}{Lin, X.} \emph{et~al.}
\newblock \bibinfo{title}{Ab initio study of electronic and optical behavior of
  two-dimensional silicon carbide}.
\newblock \emph{\bibinfo{journal}{J. Mater. Chem. C}}
  \textbf{\bibinfo{volume}{1}}, \bibinfo{pages}{2131--2135}
  (\bibinfo{year}{2013}).

\bibitem{Shi15AN}
\bibinfo{author}{Shi, Z.}, \bibinfo{author}{Zhang, Z.},
  \bibinfo{author}{Kutana, A.} \& \bibinfo{author}{Yakobson, B.~I.}
\newblock \bibinfo{title}{Predicting two-dimensional silicon carbide
  monolayers}.
\newblock \emph{\bibinfo{journal}{ACS Nano}} \textbf{\bibinfo{volume}{9}},
  \bibinfo{pages}{9802--9809} (\bibinfo{year}{2015}).

\bibitem{Lin15JPCC}
\bibinfo{author}{Lin, S.} \emph{et~al.}
\newblock \bibinfo{title}{Quasi-two-dimensional sic and sic2: Interaction of
  silicon and carbon at atomic thin lattice plane}.
\newblock \emph{\bibinfo{journal}{The Journal of Physical Chemistry C}}
  \textbf{\bibinfo{volume}{119}}, \bibinfo{pages}{19772--19779}
  (\bibinfo{year}{2015}).
\newblock \eprint{http://dx.doi.org/10.1021/acs.jpcc.5b04113}.

\bibitem{Chabi16N}
\bibinfo{author}{Chabi, S.}, \bibinfo{author}{Chang, H.}, \bibinfo{author}{Xia,
  Y.} \& \bibinfo{author}{Zhu, Y.}
\newblock \bibinfo{title}{From graphene to silicon carbide: ultrathin silicon
  carbide flakes}.
\newblock \emph{\bibinfo{journal}{Nanotechnology}}
  \textbf{\bibinfo{volume}{27}}, \bibinfo{pages}{075602}
  (\bibinfo{year}{2016}).

\bibitem{Dikin07N}
\bibinfo{author}{Dikin, D.~A.} \emph{et~al.}
\newblock \bibinfo{title}{Preparation and characterization of graphene oxide
  paper}.
\newblock \emph{\bibinfo{journal}{Nature}} \textbf{\bibinfo{volume}{448}},
  \bibinfo{pages}{457--460} (\bibinfo{year}{2007}).

\bibitem{Robertson15AN}
\bibinfo{author}{Robertson, A.~W.} \emph{et~al.}
\newblock \bibinfo{title}{Atomic structure of graphene subnanometer pores}.
\newblock \emph{\bibinfo{journal}{ACS Nano}} \textbf{\bibinfo{volume}{9}},
  \bibinfo{pages}{11599--11607} (\bibinfo{year}{2015}).

\bibitem{Lee13NC}
\bibinfo{author}{Lee, J.}, \bibinfo{author}{Zhou, W.},
  \bibinfo{author}{Pennycook, S.~J.}, \bibinfo{author}{Idrobo, J.-C.} \&
  \bibinfo{author}{Pantelides, S.~T.}
\newblock \bibinfo{title}{Direct visualization of reversible dynamics in a
  {Si}$_6$ cluster embedded in a graphene pore}.
\newblock \emph{\bibinfo{journal}{Nature}} \textbf{\bibinfo{volume}{4}},
  \bibinfo{pages}{1650} (\bibinfo{year}{2013}).

\bibitem{Skakalova14C}
\bibinfo{author}{Sk\'akalov\'a, V.} \emph{et~al.}
\newblock \bibinfo{title}{Electronic transport in composites of graphite oxide
  with carbon nanotubes}.
\newblock \emph{\bibinfo{journal}{Carbon}} \textbf{\bibinfo{volume}{72}},
  \bibinfo{pages}{224 -- 232} (\bibinfo{year}{2014}).

\bibitem{Krivanek10N}
\bibinfo{author}{Krivanek, O.~L.} \emph{et~al.}
\newblock \bibinfo{title}{Atom-by-atom structural and chemical analysis by
  annular dark-field electron microscopy}.
\newblock \emph{\bibinfo{journal}{Nature}} \textbf{\bibinfo{volume}{464}},
  \bibinfo{pages}{571--574} (\bibinfo{year}{2010}).

\bibitem{Zhao14S}
\bibinfo{author}{Zhao, J.} \emph{et~al.}
\newblock \bibinfo{title}{Free-standing single-atom-thick iron membranes
  suspended in graphene pores}.
\newblock \emph{\bibinfo{journal}{Science}} \textbf{\bibinfo{volume}{343}},
  \bibinfo{pages}{1228--1232} (\bibinfo{year}{2014}).

\bibitem{Zan12NL}
\bibinfo{author}{Zan, R.}, \bibinfo{author}{Ramasse, Q.~M.},
  \bibinfo{author}{Bangert, U.} \& \bibinfo{author}{Novoselov, K.~S.}
\newblock \bibinfo{title}{Graphene reknits its holes}.
\newblock \emph{\bibinfo{journal}{Nano Letters}} \textbf{\bibinfo{volume}{12}},
  \bibinfo{pages}{3936--3940} (\bibinfo{year}{2012}).

\bibitem{Chen16AN}
\bibinfo{author}{Chen, Q.} \emph{et~al.}
\newblock \bibinfo{title}{Elongated silicon--carbon bonds at graphene edges}.
\newblock \emph{\bibinfo{journal}{ACS Nano}} \textbf{\bibinfo{volume}{10}},
  \bibinfo{pages}{142--149} (\bibinfo{year}{2016}).

\bibitem{Enkovaara2010}
\bibinfo{author}{Enkovaara, J.} \emph{et~al.}
\newblock \bibinfo{title}{Electronic structure calculations with {GPAW}: a
  real-space implementation of the projector augmented-wave method}.
\newblock \emph{\bibinfo{journal}{J. Phys. Condens. Matter}}
  \textbf{\bibinfo{volume}{22}}, \bibinfo{pages}{253202}
  (\bibinfo{year}{2010}).

\bibitem{Koch2002}
\bibinfo{author}{Koch, C.}
\newblock \emph{\bibinfo{title}{Determination of Core Structure Periodicity and
  Point Defect Density along Dislocations}}.
\newblock Ph.D. thesis, \bibinfo{school}{Arizona State University}
  (\bibinfo{year}{2002}).

\bibitem{Kuramochi09UM}
\bibinfo{author}{Kuramochi, K.} \emph{et~al.}
\newblock \bibinfo{title}{Effect of chromatic aberration on atomic-resolved
  spherical aberration corrected {STEM} images}.
\newblock \emph{\bibinfo{journal}{Ultramicroscopy}}
  \textbf{\bibinfo{volume}{110}}, \bibinfo{pages}{36 -- 42}
  (\bibinfo{year}{2009}).

\bibitem{Forbes11UM}
\bibinfo{author}{Forbes, B.} \emph{et~al.}
\newblock \bibinfo{title}{Thermal diffuse scattering in transmission electron
  microscopy}.
\newblock \emph{\bibinfo{journal}{Ultramicroscopy}}
  \textbf{\bibinfo{volume}{111}}, \bibinfo{pages}{1670 -- 1680}
  (\bibinfo{year}{2011}).

\bibitem{huang12NL}
\bibinfo{author}{Huang, P.~Y.} \emph{et~al.}
\newblock \bibinfo{title}{Direct imaging of a two-dimensional silica glass on
  graphene}.
\newblock \emph{\bibinfo{journal}{Nano Lett.}} \textbf{\bibinfo{volume}{12}},
  \bibinfo{pages}{1081--1086} (\bibinfo{year}{2012}).

\bibitem{Tararan16CM}
\bibinfo{author}{Tararan, A.}, \bibinfo{author}{Zobelli, A.},
  \bibinfo{author}{Benito, A.~M.}, \bibinfo{author}{Maser, W.~K.} \&
  \bibinfo{author}{St{\'e}phan, O.}
\newblock \bibinfo{title}{Revisiting graphene oxide chemistry via
  spatially-resolved electron energy loss spectroscopy}.
\newblock \emph{\bibinfo{journal}{Chemistry of Materials}}
  \textbf{\bibinfo{volume}{28}}, \bibinfo{pages}{3741--3748}
  (\bibinfo{year}{2016}).

\bibitem{Susi14PRL}
\bibinfo{author}{Susi, T.} \emph{et~al.}
\newblock \bibinfo{title}{Silicon--carbon bond inversions driven by 60-ke{V}
  electrons in graphene}.
\newblock \emph{\bibinfo{journal}{Phys. Rev. Lett.}}
  \textbf{\bibinfo{volume}{113}}, \bibinfo{pages}{115501}
  (\bibinfo{year}{2014}).

\bibitem{Guan16NL}
\bibinfo{author}{Guan, J.}, \bibinfo{author}{Liu, D.}, \bibinfo{author}{Zhu,
  Z.} \& \bibinfo{author}{Tom{\'a}nek, D.}
\newblock \bibinfo{title}{Two-dimensional phosphorus carbide: Competition
  between sp$^2$ and sp$^3$ bonding}.
\newblock \emph{\bibinfo{journal}{Nano Letters}} \textbf{\bibinfo{volume}{16}},
  \bibinfo{pages}{3247--3252} (\bibinfo{year}{2016}).

\bibitem{Tang09JoPCM}
\bibinfo{author}{Tang, W.}, \bibinfo{author}{Sanville, E.} \&
  \bibinfo{author}{Henkelman, G.}
\newblock \bibinfo{title}{A grid-based bader analysis algorithm without lattice
  bias}.
\newblock \emph{\bibinfo{journal}{J. Phys.: Condens. Matter}}
  \textbf{\bibinfo{volume}{21}}, \bibinfo{pages}{084204}
  (\bibinfo{year}{2009}).

\bibitem{Thygesen05PRL}
\bibinfo{author}{Thygesen, K.~S.}, \bibinfo{author}{Hansen, L.~B.} \&
  \bibinfo{author}{Jacobsen, K.~W.}
\newblock \bibinfo{title}{Partly occupied {W}annier functions}.
\newblock \emph{\bibinfo{journal}{Phys. Rev. Lett.}}
  \textbf{\bibinfo{volume}{94}}, \bibinfo{pages}{026405}
  (\bibinfo{year}{2005}).

\bibitem{Kackell94PRB}
\bibinfo{author}{K\"ackell, B., P.;~Wenzien} \& \bibinfo{author}{Bechstedt, F.}
\newblock \bibinfo{title}{Influence of atomic relaxations on the structural
  properties of {SiC} polytypes from \textit{ab initio} calculations}.
\newblock \emph{\bibinfo{journal}{Phys. Rev. B}} \textbf{\bibinfo{volume}{50}},
  \bibinfo{pages}{17037--17046} (\bibinfo{year}{1994}).

\bibitem{Kotakoski11PRB}
\bibinfo{author}{Kotakoski, J.} \emph{et~al.}
\newblock \bibinfo{title}{Stone-{W}ales-type transformations in carbon
  nanostructures driven by electron irradiation}.
\newblock \emph{\bibinfo{journal}{Phys. Rev. B}} \textbf{\bibinfo{volume}{83}},
  \bibinfo{pages}{245420} (\bibinfo{year}{2011}).

\bibitem{Susi12AN}
\bibinfo{author}{Susi, T.} \emph{et~al.}
\newblock \bibinfo{title}{Atomistic description of electron beam damage in
  nitrogen-doped graphene and single-walled carbon nanotubes}.
\newblock \emph{\bibinfo{journal}{ACS Nano}} \textbf{\bibinfo{volume}{6}},
  \bibinfo{pages}{8837--8846} (\bibinfo{year}{2012}).

\bibitem{Kotakoski12AN}
\bibinfo{author}{Kotakoski, J.}, \bibinfo{author}{Santos-Cottin, D.} \&
  \bibinfo{author}{Krasheninnikov, A.~V.}
\newblock \bibinfo{title}{Stability of graphene edges under electron beam:
  Equilibrium energetics versus dynamic effects}.
\newblock \emph{\bibinfo{journal}{ACS Nano}} \textbf{\bibinfo{volume}{6}},
  \bibinfo{pages}{671--676} (\bibinfo{year}{2012}).

\bibitem{Susi16NC}
\bibinfo{author}{Susi, T.} \emph{et~al.}
\newblock \bibinfo{title}{Isotope analysis in the transmission electron
  microscope}.
\newblock \emph{\bibinfo{journal}{Nature Communications}}
  \textbf{\bibinfo{volume}{7}}, \bibinfo{pages}{13040} (\bibinfo{year}{2016}).

\bibitem{Meyer12PRL}
\bibinfo{author}{Meyer, J.~C.} \emph{et~al.}
\newblock \bibinfo{title}{Accurate measurement of electron beam induced
  displacement cross sections for single-layer graphene}.
\newblock \emph{\bibinfo{journal}{Phys. Rev. Lett.}}
  \textbf{\bibinfo{volume}{108}}, \bibinfo{pages}{196102}
  (\bibinfo{year}{2012}).

\bibitem{Ljungberg11JESRP}
\bibinfo{author}{Ljungberg, M.~P.}, \bibinfo{author}{Mortensen, J.~J.} \&
  \bibinfo{author}{Pettersson, L. G.~M.}
\newblock \bibinfo{title}{An implementation of core level spectroscopies in a
  real space projector augmented wave density functional theory code}.
\newblock \emph{\bibinfo{journal}{J. Electron Spectros. Related Phenom.}}
  \textbf{\bibinfo{volume}{184}}, \bibinfo{pages}{427--439}
  (\bibinfo{year}{2011}).

\bibitem{Susi15PRB}
\bibinfo{author}{Susi, T.}, \bibinfo{author}{Mowbray, D.~J.},
  \bibinfo{author}{Ljungberg, M.~P.} \& \bibinfo{author}{Ayala, P.}
\newblock \bibinfo{title}{Calculation of the graphene {C} 1\textit{s} core
  level binding energy}.
\newblock \emph{\bibinfo{journal}{Phys. Rev. B}} \textbf{\bibinfo{volume}{91}},
  \bibinfo{pages}{081401} (\bibinfo{year}{2015}).

\bibitem{Sofo07PRB}
\bibinfo{author}{Sofo, J.~O.}, \bibinfo{author}{Chaudhari, A.~S.} \&
  \bibinfo{author}{Barber, G.~D.}
\newblock \bibinfo{title}{Graphane: A two-dimensional hydrocarbon}.
\newblock \emph{\bibinfo{journal}{Phys. Rev. B}} \textbf{\bibinfo{volume}{75}},
  \bibinfo{pages}{153401} (\bibinfo{year}{2007}).

\bibitem{Molle13AFM}
\bibinfo{author}{Molle, A.} \emph{et~al.}
\newblock \bibinfo{title}{Hindering the oxidation of silicene with non-reactive
  encapsulation}.
\newblock \emph{\bibinfo{journal}{Advanced Functional Materials}}
  \textbf{\bibinfo{volume}{23}}, \bibinfo{pages}{4340--4344}
  (\bibinfo{year}{2013}).

\bibitem{Wood14NL}
\bibinfo{author}{Wood, J.~D.} \emph{et~al.}
\newblock \bibinfo{title}{Effective passivation of exfoliated black phosphorus
  transistors against ambient degradation}.
\newblock \emph{\bibinfo{journal}{Nano Letters}} \textbf{\bibinfo{volume}{14}},
  \bibinfo{pages}{6964--6970} (\bibinfo{year}{2014}).

\bibitem{Cooper10PRB}
\bibinfo{author}{Cooper, V.~R.}
\newblock \bibinfo{title}{Van der {W}aals density functional: An appropriate
  exchange functional}.
\newblock \emph{\bibinfo{journal}{Phys. Rev. B}} \textbf{\bibinfo{volume}{81}},
  \bibinfo{pages}{161104} (\bibinfo{year}{2010}).

\bibitem{Constantinescu13PRL}
\bibinfo{author}{Constantinescu, G.}, \bibinfo{author}{Kuc, A.} \&
  \bibinfo{author}{Heine, T.}
\newblock \bibinfo{title}{Stacking in bulk and bilayer hexagonal boron
  nitride}.
\newblock \emph{\bibinfo{journal}{Phys. Rev. Lett.}}
  \textbf{\bibinfo{volume}{111}}, \bibinfo{pages}{036104}
  (\bibinfo{year}{2013}).

\bibitem{Padilha15JPCC}
\bibinfo{author}{Padilha, J.~E.} \& \bibinfo{author}{Pontes, R.~B.}
\newblock \bibinfo{title}{Free-standing bilayer silicene: The effect of
  stacking order on the structural, electronic, and transport properties}.
\newblock \emph{\bibinfo{journal}{The Journal of Physical Chemistry C}}
  \textbf{\bibinfo{volume}{119}}, \bibinfo{pages}{3818--3825}
  (\bibinfo{year}{2015}).

\bibitem{Geim13N}
\bibinfo{author}{Geim, A.~K.} \& \bibinfo{author}{Grigorieva, I.~V.}
\newblock \bibinfo{title}{Van der {W}aals heterostructures}.
\newblock \emph{\bibinfo{journal}{Nature}} \textbf{\bibinfo{volume}{499}},
  \bibinfo{pages}{419--425} (\bibinfo{year}{2013}).

\bibitem{Meyer08NL}
\bibinfo{author}{Meyer, J.~C.} \emph{et~al.}
\newblock \bibinfo{title}{Direct imaging of lattice atoms and topological
  defects in graphene membranes}.
\newblock \emph{\bibinfo{journal}{Nano Lett.}} \textbf{\bibinfo{volume}{8}},
  \bibinfo{pages}{3582--3586} (\bibinfo{year}{2008}).

\bibitem{IoffeSiC}
\bibinfo{author}{{Ioffe Institute}}.
\newblock \bibinfo{title}{Properties of silicon carbide.}
\newblock \urlprefix\url{http://www.ioffe.ru/SVA/NSM/Semicond/SiC/}.

\bibitem{Binner01JMSL}
\bibinfo{author}{Binner, J.} \& \bibinfo{author}{Zhang, Y.}
\newblock \bibinfo{title}{Characterization of silicon carbide and silicon
  powders by {XPS} and zeta potential measurement}.
\newblock \emph{\bibinfo{journal}{Journal of Materials Science Letters}}
  \textbf{\bibinfo{volume}{20}}, \bibinfo{pages}{123--126}
  (\bibinfo{year}{2001}).

\bibitem{Johansson96SS}
\bibinfo{author}{Johansson, L.}, \bibinfo{author}{Owman, F.} \&
  \bibinfo{author}{M{\aa}rtensson, P.}
\newblock \bibinfo{title}{A photoemission study of {4HSiC}(0001)}.
\newblock \emph{\bibinfo{journal}{Surface Science}}
  \textbf{\bibinfo{volume}{360}}, \bibinfo{pages}{483 -- 488}
  (\bibinfo{year}{1996}).

\bibitem{Parrill91SS}
\bibinfo{author}{Parrill, T.~M.} \& \bibinfo{author}{Chung, Y.~W.}
\newblock \bibinfo{title}{{Surface analysis of cubic silicon carbide (001)}}.
\newblock \emph{\bibinfo{journal}{Surface Science}}
  \textbf{\bibinfo{volume}{243}}, \bibinfo{pages}{96--112}
  (\bibinfo{year}{1991}).

\end{thebibliography}
\end{document}